# Diagnostics Examples from Third-Generation Light Sources

*Kay Wittenburg*
Deutsches Elektronen Synchrotron, DESY, Hamburg, Germany

**Abstract**
This lesson discusses many examples of how the signals from the beam monitors are used to diagnose the beam in circular, third-generation synchrotron light sources. During the school, diagnostic examples in other machines (e.g. colliders, CTF3, linacs and free-electron lasers (FEL), and medical accelerators) were given in other lectures. This lesson assumes that the signal generation in the instrument itself is already known; the main focus lies on the dependence of the signals on various machine parameters and their interpretation to diagnose the machine parameters and conditions.

## 1 Introduction

### 1.1 What is special about third-generation light sources?

Third-generation light sources deliver high brilliance X-ray beams to the experimental stations. The brilliance (or spectral brightness) $B_{\text{ph}}$ of the X-ray photon beam (ph) is defined by

$$B_{\text{ph}} = \frac{N_{\text{ph}}}{\Delta t \cdot \varepsilon_{\text{ph},x} \cdot \varepsilon_{\text{ph},y} \cdot 0.1\% BW} \left[\frac{\text{Photons}}{s \cdot \text{mm}^2 \cdot \text{mrad}^2 \cdot 0.1\% BW}\right],$$

where $N_{\text{ph}}/0.1\% BW$ is the number of photons within the bandwidth (wavelength interval of interest), $\Delta t$ is the time interval, and $\varepsilon_{\text{ph}}$ is the emittance of the photon beam. The shortest time interval is defined by the bunch length $\sigma_t$ [$s$], and the emittance is related to the beam size and divergence (normalized beam emittance $\varepsilon_n$). The bandwidth of the photon beam is related to the insertion device (ID) properties, but also to the beam energy and energy spread $\sigma_\gamma \sim \Delta p/p$. The 6D beam brightness at a given beam energy is hence defined (see Ref. [1]) by

$$B_{6\text{D}} \sim \frac{Ne}{\varepsilon_{n,x} \cdot \varepsilon_{n,y} \cdot \sigma_t \cdot \sigma_\gamma} \left[\frac{A}{m^2}\right],$$

where $Ne$ = number of electrons in a time interval $\sigma_t$ (current, $I$). These formulas already imply most of the important parameters to be measured by beam instrumentation and derived by diagnostics to achieve a high brightness beam, which is the goal of each third-generation synchrotron light source: very small *emittances* and *small β-function values* to achieve very small *beam sizes* and *angular spread* in both planes $x$ and $y$ in the IDs; typical size values are 100–300 μm in the horizontal plane and below 10 μm in the vertical plane; typical emittance values are 10 prad (vertically) and 1 nrad (horizontally) . Since each of the experimental stations need very stable beam conditions (to keep the small emittance), the *beam orbit* has to be stabilized to about 10% of the beam size during the experiment time, and therefore over many days. All kind of mechanical (e.g. temperature) and electronic (beam position monitors, BPM) drifts have to be minimized, while a very good *orbit feedback* system is required to remove residual orbit distortions (see Ref. [2]). The small emittances of the beam in both planes imply a very small *coupling* of the two planes and a well understood beam optics, e.g. *tune*, *dispersion function D* and *chromaticity*. Since the experiments ask for high intensities, the light source has to work with *multi-bunch fillings* with high bunch charge ($\sigma_t$ = bunch length) and/or high average currents (DC current in the ring, $\sigma_t$ = revolution time), which forces many kinds of instabilities so that a *multi-bunch feedback* is often required. Many third-generation light sources are working in a top-up mode, which ensures a very stable *(DC) beam current* and stable *(AC) bunch current* in the machine to avoid high temperature variations on machine components. A comprehensive treatment of the physics of low emittance storage rings can be found in Ref. [3]. Table 1 summarises the basic machine parameter of some third-generation light sources; more can be found in Ref. [4].

**Table 1:** Parameters of some third-generation light sources

| Facility | Year | Energy [GeV] | Current [mA] | Circumference [m] | Emittance $\varepsilon_x$ [nm rad] | Emittance $\varepsilon_y$ [pm rad] |
|---|---|---|---|---|---|---|
| SLS | 2001 | 2.4 | 400 | 288 | 5 | 35 |
| CLS | 2001 | 2.9 | 500 | 171 | 20 | 92 |
| Soleil | 2007 | 2.75 | 500 | 354 | 3.7 | 37 |
| Diamond | 2007 | 3 | 300 | 562 | 2.7 | 27 |
| SSRF | 2008 | 3.5 | 300 | 432 | 3.9 | 39 |
| PETRA III | 2009 | 6 | 100 | 2304 | 1.2 | 1.5 |
| ALBA | 2010 | 3 | 400 | 269 | 4.3 | 40 |
| ESRF-U | 2011 | 6 | 300 | 844 | 4 | 10 |
| NSLS-II | 2015 | 3 | 500 | 792 | 0.9 | 8 |
| MAX IV | 2015 | 3 | 500 | 582 | 0.3 | 6 |

**1.2 List of diagnostic examples discussed in this report**

The following diagnostic examples are discussed:

a. Beam position monitors and beam stabilization;
b. Orbit concepts: beam-based alignment, corrections, 'golden orbit', etc.;
c. Optic functions: beta-function, dispersion, tune, coupling, chromaticity, centre frequency;
d. Stabilizing the beam: feedbacks, instabilities;
e. Beam current: top-up, transfer efficiency, lifetime (dynamic aperture, Touschek, etc.);
f. Beam size: emittance;
g. Bunch length: energy spread, instabilities, bunch purity;
h. Beam losses: radiation damage of undulators.

Items a–d are machine parameters that are typically measured by BPMs, which makes the BPMs one of the most important instrument in such machines. Therefore, some relevant parameters of the BPM readout are discussed in Chapter 2. The order of the remaining topics approximately follows the order of topics for commissioning a new machine. Far more detailed views on the different instruments (BPM, current, size, length, etc.) are given in various lessons of these CERN Accelerator School proceedings. Many of the following examples came from the commissioning of the PETRA III synchrotron light source. Most of the pictures are copied from the PETRA Logbook [5], illustrating real measurements during commissioning and machine studies.

Detailed discussions of the various examples of experiments and measurements to diagnose the beam cannot be given in this report, since each single topic often needs a lengthy explanation. These details have to be read about in the given references. I have tried to refer in general to more modern reports, while older ones can often be found in the reference lists of the reports listed.

## 2  Beam position monitors and their stabilization

A typical synchrotron light source has different beam pipe diameters and shapes, e.g. in the arcs, near and within the IDs, at the damping wigglers, at injection, etc. Therefore, various BPM geometries exist in the same ring; a small monitor constant of a BPM geometry is very helpful for achieving a good beam position resolution. The best resolution of the BPMs is required at the undulators to give the orbit feedback system (see Chapter 5) the best chance of keeping the beam as stable as possible at these locations. A resolution of about one-tenth of the beam size $\sigma$ in both planes is a requirement of the synchrotron light's users to keep the light intensity on their target quite stable. This results in a closed orbit resolution requirement below 1 μm, typically in the vertical plane. Modern BPM electronic readouts reach 0.3 μm [6, 7] averaged over many turns. To keep the readout stable at such a level, active stabilization in the BPM readout chain is required, see Chapter 2.1–2.3. Passive stabilization methods like massive concrete foundations and rigid girders are, however, very helpful for keeping accelerator and experiment component vibration as small as possible.

### 2.1  Electronic stabilization

Different drift behaviours in the four readout channels of a BPM lead to (fake) drifts in the beam position, even if the beam does not move. An orbit feedback system will, however, note the (fake) change of the beam position and correct it, with the result that the beam now moves while the readout beam position is kept constant. This kind of electronic drift has been cancelled in many modern BPM readout systems by various methods. Reference [7] uses an analogue switching crossbar with an internal learning algorithm for keeping each channel calibrated, while Ref. [8] uses a pilot tone concept for dynamic calibration of each channel.

### 2.2  Temperature stabilization

Even with electronic stabilization, electronic drift with temperature remains on a level of some 0.1 μm/°C [7]. Therefore, temperature stabilization of the readout electronics in a range better than 1°C is required. This might require acclimatized closed racks or a housing for the racks with temperature stabilization in this range.

An effect of air humidity on the parameters of the (typically very good quality) cables between the BPM and the readout electronic was reported [9], which can influence the beam position readout as well. So far, this parameter is hardly to be controlled, but for extreme stabilization requirements for future light sources it might become important.

### 2.3  Mechanical movement (drift) of beam pipe

The BPMs are typically fix-points in the vacuum system of almost all accelerators, since any kind of movement of a BPM will immediately be shown as a (fake) change in the beam position. BPMs are often almost decoupled from the other parts of the vacuum system by bellows, but this isn't possible at all locations. In synchrotron light sources the emitted synchrotron light from the bending magnets hits the outer beam pipe and deposits a lot of power onto it. Therefore, water cooling is required on the side to avoid overheating of the material. But there is always a temperature gradient across the bended beam pipe, which implies quite a force on the whole segment. This can be so strong that the beam pipe can touch the quadrupoles, and can even move the magnets [10], with an unwanted effect on the beam orbit. Nowadays, light sources use girder concepts to avoid mechanical movements and vibrations [11], together with decoupled beam pipes from the magnets, and even temperature-stabilized accelerator tunnels. The temperature gradient across the vacuum pipe with stored beam is, however, still present. Measurements at PETRA III with movement monitors on various BPMs [12] have shown a movement of BPMs of many tens of μm between zero and design beam current (100 mA), see Fig. 1. Therefore, a constant beam current (e.g. via top-up) is most helpful for avoiding this kind of BPM movement. A re-stabilization time of some minutes after beam injection is also required to counterbalance all remaining movements.

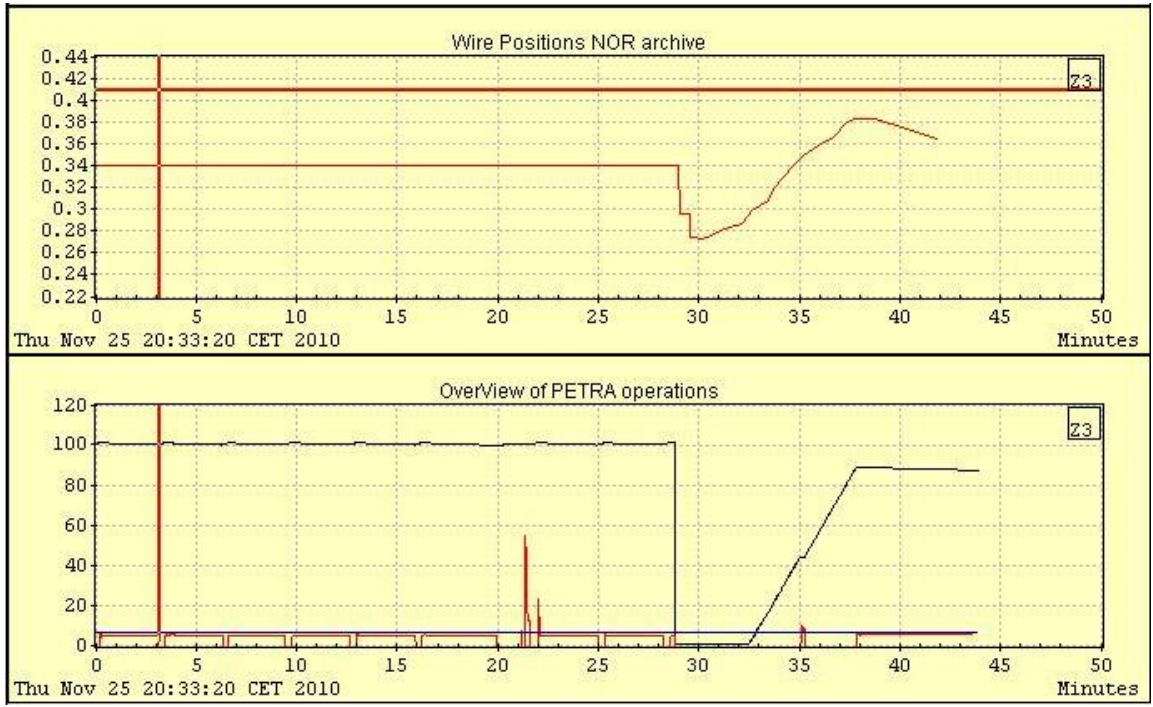

**Fig. 1:** Lower part: Beam current drop from 100 mA to 0 mA and adjacent injection to 90 mA. Upper part: Response of a movement monitor connected to a BPM. The BPM moves in the horizontal plane by more than ±60 μm due to the beam current drop and current recovery [5].

## 3    Orbit concepts: Beam based alignment, corrections, 'golden orbit'

### 3.1    Beam-based alignment

At the first commissioning of any accelerator one has to rely on the alignment measurements of the BPMs and the magnets within the accelerator tunnel. While storing the very first beam, its orbit might show offsets of some or even many millimetres. At this moment it is typically not clear whether the BPM itself has an alignment or electronic offset, or if the beam is really out of the BPM centre. Since the quadrupoles of the storage ring are defining the optic parameters of the ring, it is desirable that the beam will go through the centre of the quadrupoles. For this reason, it is also desirable to have BPMs very close to all quadrupoles, and that the magnetic centre of a quadrupole defines the zero reading of the adjacent BPM. The procedure to find the offset of the BPM with respect to the centre of the quadrupole is called beam-based alignment [13, 14].

An off-centre beam in a quadrupole results in an additional dipole kick, and therefore in betatron oscillations around the ring, which can be measured by all BPMs by difference-orbits. The amplitude of the betatron oscillation depends on the quadrupole excitation (magnet current), while in the centre of the quadrupole the beam will not receive a dipole kick, independent of the quadrupole excitation. The beam is then moved with the help of local orbit bumps across many positions inside one quadrupole, while for each position the quadrupole excitation is changed, and the global orbit is measured in parallel by all BPMs. At a local beam position where the betatron oscillation amplitude is independent from the quadrupole excitation, the beam is in the centre of the quadrupole, and the reading of the adjacent BPM is defined as the offset of the BPM. This (time-consuming) procedure has to be done for both planes and for all quadrupole–BPM pairs of the storage ring to define all BPM offsets.

## 3.2 Local orbit correction

Large orbit excursions or wanted orbit offsets (e.g. at injection) have to be corrected or adjusted to ensure a maximum aperture in the ring. Additionally, the orbit setting is particularly important at some discrete locations like IDs. An orbit correction can be aimed at suppressing the orbit distortion only at these locations using a closed bump, leaving the rest of the machine uncorrected. Such a scheme requires a minimum of one BPM for the orbit distortion and a minimum of two correctors for the local cancellation of position, angle and the bump closing (for a phase advance of 180°; more correctors are required if the phase advance is less or more than 180°).

## 3.3 'Golden orbit'

After taken into account all BPM offsets and all necessary local orbit corrections and further optimizations like injection efficiency, lifetime and aperture issues, etc., an optimum orbit can be defined. This is often called the 'golden orbit', although it might not always be the one that is centred at all quadrupoles (corrected zero of all the BPMs). Depending on the required machine parameters like bunch frequency, stored current, beam energy, optimized injection, optimized beam for particular experiments, etc., the best orbit might be different, therefore several golden orbits might exist. Further machine studies and optimizations (see the following chapters) need a base to start with. This is the respective golden orbit. After each successful optimization, however, the recent orbit might become a new golden one.

# 4 Orbit functions

## 4.1 Beta function

Assume in the following that all BPM readings for a golden orbit are defined as zero. By stimulating a beam oscillation with a short kicker pulse the beam will undergo betatron oscillations. These oscillations can be measured by all BPMs on successive non-degraded turns (turn-by-turn, TBT) after the excitation. The envelope of the measured positions follows the square root of the beta function. A fit of the data delivers the betatron amplitude (or β-function) and the corresponding phase at each BPM.

The β–function at each quadrupole (not at each BPM like above!) can be measured by modulating the current in that quadrupole and measuring the change in the betatron tune Q [15, 16]

$$\beta = 4\pi \frac{\Delta Q}{L \Delta K} \ .$$

The calibration of the quadrupole strength $K$ depends on the magnet current, and hysteresis effects can degrade the measurement.

The measured values of both methods can be compared with the theoretical values of the lattice. The resulting differences of the β-function are often called 'beta-beat'. Beta-beats smaller than 1% are typical values for synchrotron light sources.

### 4.1.1 Orbit response matrix

As well as a number of BPMs *M* in a storage ring, there is an almost equivalent number of orbit correctors *N*, typically *M* > *N*. Orbit displacements arising from corrector kicks are determined by the $M \times N$ linear orbit response matrix (ORM). The elements of the ORM can be determined by measuring the response of each BPM on the excitation of each corrector. The size of the ORM easily reaches many more than 10,000 elements, depending on the number of correctors and BPMs. The measured elements can be fitted and compared with the theoretical values to find optics errors of the real machine. One example of a 'standard' tool for synchrotron light sources is Linear Optics from Closed Orbits (LOCO) [17]. Beta-beating, phase mismatches and coupling can be obtained by this tool as well as detection of faulty BPMs [18].

## 4.2 Dispersion

Almost zero dispersion is required in all the IDs to keep the emittance of the beam as small as possible [3] (Fig. 2). The amplitude of the dispersion function $D_{x,y}$ at each BPM can be measured by the orbit difference Δx or Δy at different beam energy (or momentum *p*)

$$\Delta x = \Delta D_x \frac{\Delta p}{p_0}, \; \Delta y = \Delta D_y \frac{\Delta p}{p_0}.$$

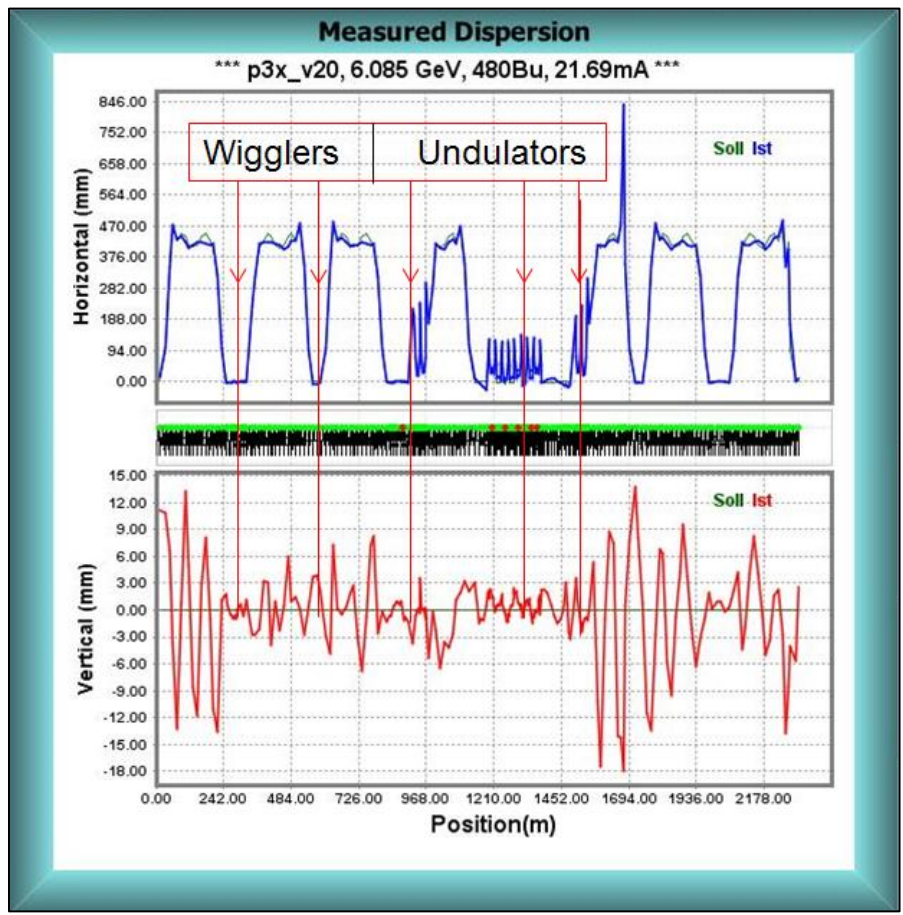

**Fig. 2:** Horizontal and vertical dispersion measurement at PETRA III [5]. The vertical dispersion is much smaller than the horizontal. The zero dispersion regions at the damping wigglers and at the undulators are clearly visible. The theoretical dispersion line (soll) is almost hidden by the measured one (ist), indicating small errors of the dispersion function alone.

The beam momentum change is made by varying the RF frequency $\omega_{RF}$ with constant magnets

$$\frac{\Delta\omega_{RF}}{\omega_{RF}} = \eta_c \frac{\Delta p}{p_0}$$

where $\eta_c = \alpha_c - 1/\gamma^2$ and $\alpha_c$ is the momentum compaction factor.

Only small changes are allowed depending on the momentum acceptance of the storage ring.

**4.3  Tune**

The tune of a synchrotron is the characteristic frequency of a single particle (incoherent tune) or a bunch of particles (coherent tune) around its nominal orbit. The two tunes might differ slightly because of various effects, e.g. space charge, beam–beam effects or others, but these are almost negligible in third-generation light sources (see also Chapter 8.1). In electron synchrotrons there is hardly any access to the incoherent tune; therefore, the main interest lies in a value for the coherent tune. The incoherent tune amplitudes, however, define the beam size of the stored beam. Both tunes are given by the magnet lattice, especially by the strength of the quadrupole magnets. In these proceedings one can find a dedicated session about tune and chromaticity measurements [19], therefore no special details will be given here.

Within one turn around the machine there are many oscillations around the nominal orbit: these can be divided in the integer tune $Q_{int}$ ($2\pi$ oscillations) and the fractional tune $q_{frac}$ (remaining part the last oscillation) with $Q = Q_{int} + q_{frac}$. During the commissioning of a ring the very first proof of the lattice is to measure and verify the expected integer tune. This is typically done by an off-axis injection of one (or more) bunches or a fast kick of a stored bunch while observing their beam positions on their first turn. By subtracting the positions of the stored, unperturbed beam (nominal orbit), the remaining full cycle ($2\pi$) oscillations around the ring can be counted, which gives $Q_{int}$ of the measured transverse plane. Depending on the size of the ring the integer tune is in the rough order of $10 < Q_{int} < 100$. A difference of the measured and expected (theoretical) $Q_{int}$ ($x,y$) indicates a large optic error in the lattice that has to be corrected immediately (e.g. a quadrupole with wrong polarity). For such a measurement one needs at least 2 BPMs per full oscillation; most rings, however, have four (or more).

Once $Q_{int}$ is proven to be correct, the main interest lies on the fractional tune $q_{frac}$ ($x,y$), in the following, named simply q. Typical tune measurements can be done by a fast kick (<1 turn) of a bunch while observing its position (betatron) oscillation at one position in the ring after this kick. This special location is often equipped with a very sensitive BPM system (e.g. Ref. [20]). Reference [21] shows a typical example of a tune setup in a light source. The time domain signals from a BPM are usually converted into the frequency domain, which presents which frequencies are present in this oscillation. Obviously, the revolution frequency is the main frequency, but the amplitude oscillation of the beam position appears as side bands of the revolution frequency (multiple) lines. By heavy filtering of the signal alone the tune frequency can be displayed.

A tune measurement via external excitation under feedback operation (see Chapter 5) is difficult because of the strong damping of the oscillations due to the feedback system. An alternative way was found to measure the tune with feedback: Looking at the signal just behind the analogue detector of the feedback system it can be seen that at the tune resonance frequency a notch will appear in the noise spectrum due to the 180° phase shift of the feedback. These notches can be analyzed, even with running feedbacks and with a minimum of excitation [22] (Fig. 3).

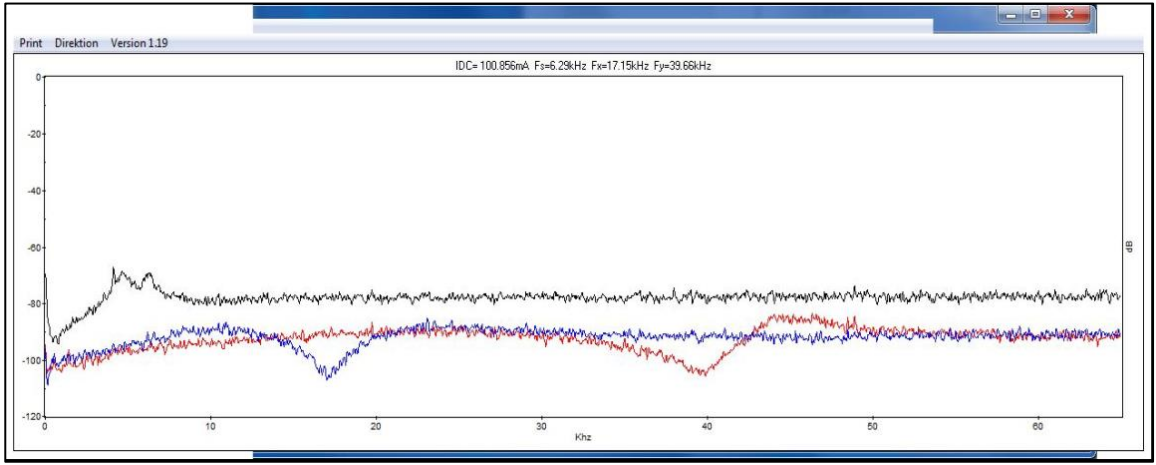

**Fig. 3:** Tune signal at PETRA III measured behind the feedback detector [5]. The dips in the noise spectrum at the tunes are clearly visible. Red: vertical tune; blue: horizontal tune; black: synchrotron tune.

### 4.4 Chromaticity

The chromaticity Q′ is defined as the change of the tune q due to a change of the momentum

$$Q' = \frac{dq}{\left(\frac{dp}{p}\right)} ,$$

where $dp/p = 1/\eta_c \, (d\omega_{RF}/\omega_{RF})$, $\eta_c$ is the slipping factor $\alpha_c - 1/\gamma^2$ and $\alpha_c$ is the momentum compaction factor.

Thus, one can measure the chromaticity by measuring the tunes as a function of the RF frequency $\omega_{RF}$. Reference [16] shows some more ways to measure chromaticity, but due to the short bunch length in synchrotron light sources (≤ a few tens of ps) a head–tail measurement is not possible. The natural chromaticity of a lattice is defined as the chromaticity with all sextupoles off, while the corrected chromaticity is measured with the sextupoles powered. The natural chromaticity is sometimes also obtained by varying the beam momentum through a dipole field change but keeping the beam on an orbit going through the sextupole centres ($dp/p = dB/B$).

Figure 4 shows examples of measurements of different chromaticities due to various settings of the sextupoles in PETRA III. The nonlinearities in the curves are the results of second-order sextupoles' cross-talk terms [23].

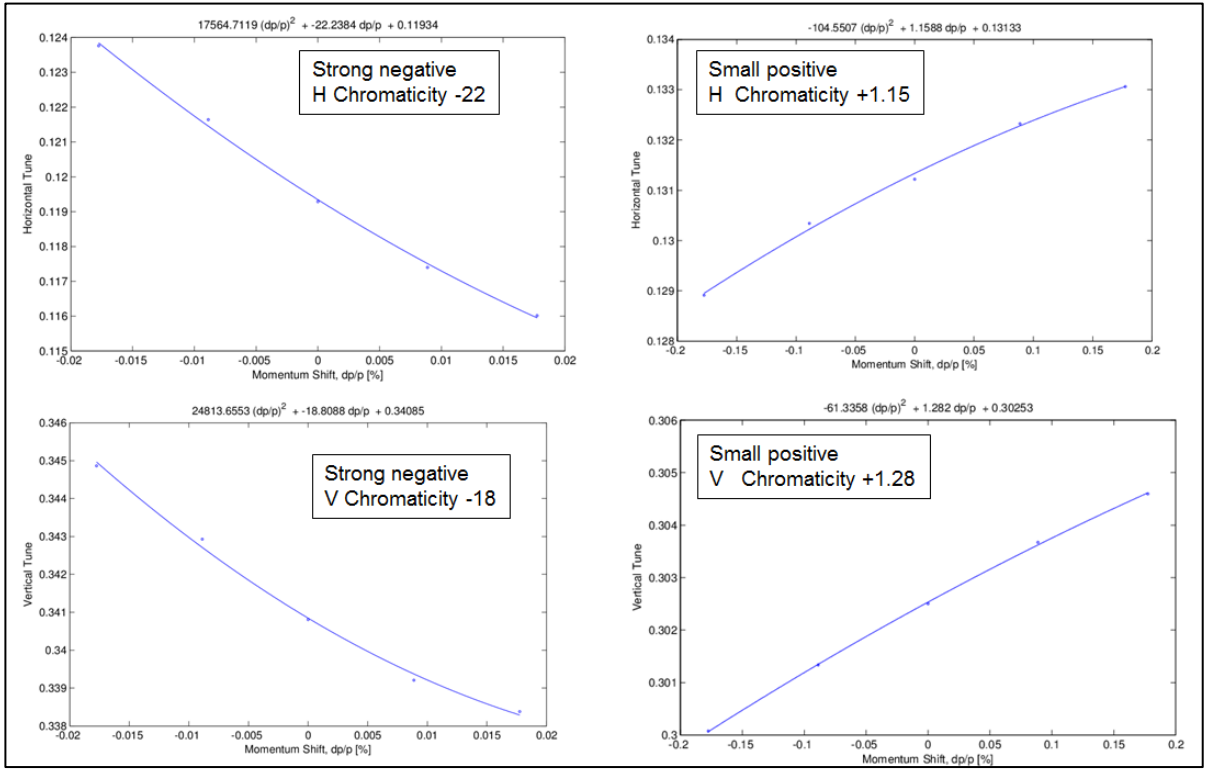

**Fig. 4:** Chromaticity measurements (vertical V and horizontal H) at PETRA III by analyzing the tune change with beam momentum [5]. The upper part shows the horizontal chromaticity, left with strong negative chromaticity of $Q'_h = -22$; and right with a small chromaticity of $Q'_h = +1.15$. The lower part shows the vertical chromaticity with $Q'_v = -18$ (left); and $Q'_v = +1.28$.

Typically, synchrotron light sources like to run with small positive chromaticity, while negative chromaticity drives instabilities (beam losses). Figure 5 shows a scan of the chromaticity versus the maximum stored bunch current Ib at PETRA III. At small negative chromaticity the bunch current was limited by head–tail instabilities, while at very large negative chromaticity the injection efficiency became very poor, probably because of a limited dynamic aperture.

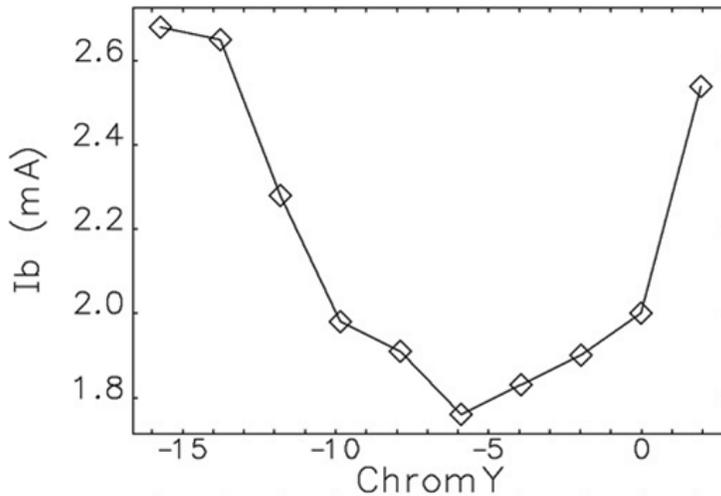

**Fig. 5:** Vertical chromaticity versus maximum stored bunch current at PETRA III [5]

## 4.5 Centre frequency

By plotting the (horizontal) tune–momentum dependencies for different chromaticities, all lines will cross each other at a certain orbit circumference where the beam goes through the mean centre of all sextupoles (and with a good approximation through the centres of the quadrupoles) (Fig. 6). This can be defined as the real circumference of the lattice, and the momentum of the electrons should be adjusted to this circumference. Actually, this point defines the right RF (centre) frequency that adjusts the orbit to the circumference (at a given dipole field).

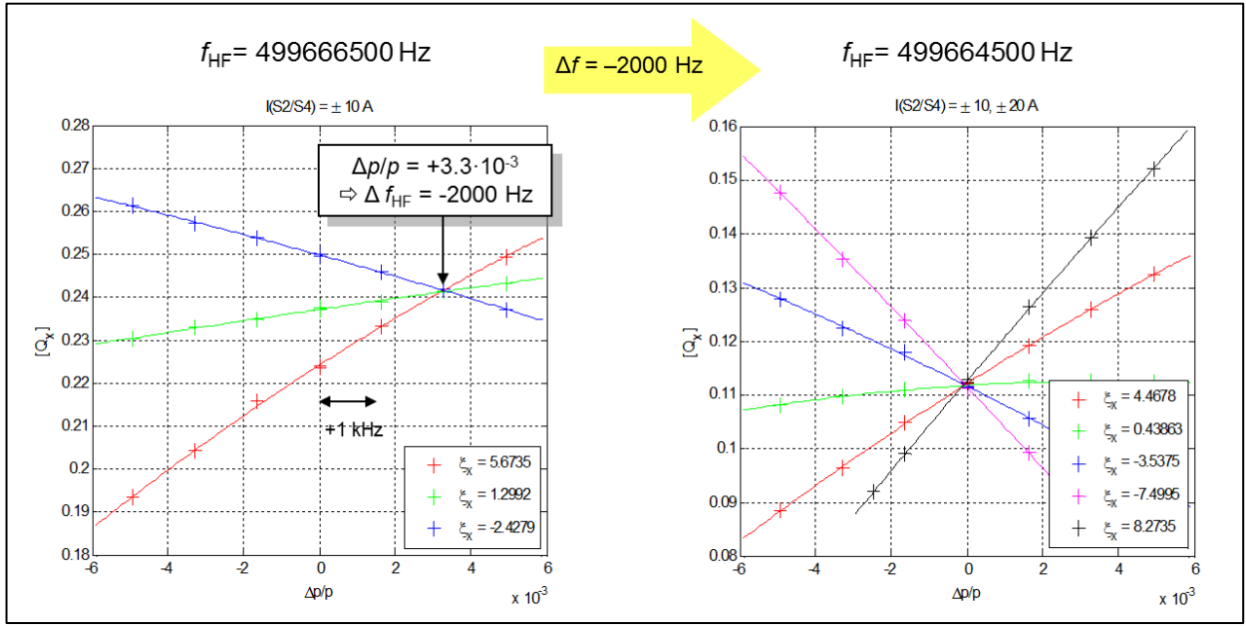

**Fig. 6:** Measurement and correction of the centre frequency of PETRA III to the circumference (= 2303.952 m) [14]. The tune $Q_x$ is measured versus momentum $\Delta p/p$ at different chromaticities $\xi_x$.

## 4.6 Coupling

The horizontal and vertical betatron oscillations are not obviously independent. Slightly rotated (misaligned) quadrupoles cannot be avoided completely; therefore in a typical lattice some coupling between the oscillations is present. In synchrotron light sources the oscillations should be almost decoupled to avoid an emittance exchange between the two planes and to achieve a very small vertical emittance. The coupling can be corrected by powering special rotated 'skew' quadrupoles in the lattice. Reference [19] discuss various ways to measure the coupling. A very common way is to use the 'closest tune approach', simply by measuring both tunes by varying the quadrupole strength. The coupling is then derived from how close the tunes can approach. Figure 7 shows a measurement of the coupling during the commissioning of PETRA III, which illustrates the principle of such a measurement. Using a skew-quad correction predicted by LOCO [24] achieved a very small tune separation of $\Delta Q = 0.0004$, and hence 0.002% betatron coupling and a measured emittance ratio of 0.17%. Since the vertical beam size in synchrotron light sources is mainly defined by the residual coupling, the measurement of the vertical beam size and the precise orientation of the beam ellipse is another way to analyze the coupling. Such dependence is shown in Ref. [24], while a very precise beam profile monitor is required to resolve typical vertical beam sizes below 20 μm as well as small tilts in the beam ellipse orientation.

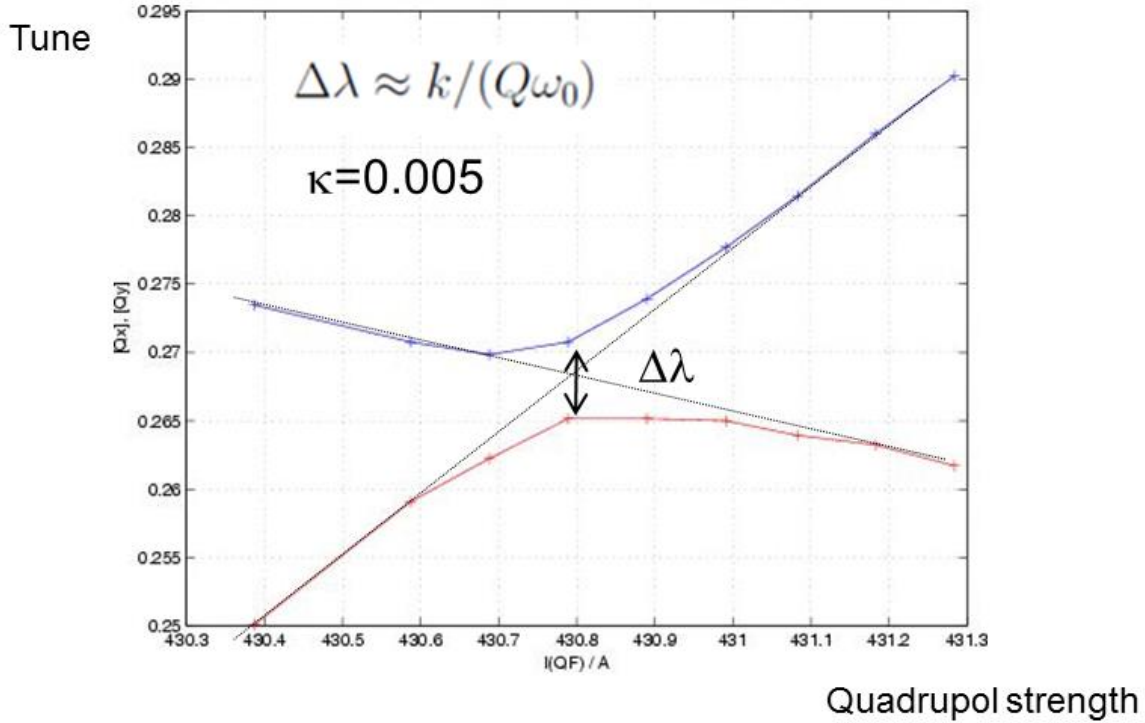

**Fig. 7:** Coupling measurement of an uncorrected lattice during commissioning of PETRA III [5] with coupling constant $k = 0.005 \approx \Delta\lambda \times (Q\omega)$. The corrected lattice reaches a coupling of better than 0.1%, but the larger distance of the tunes in an uncorrected machine illustrates better the principle of this measurement.

## 5    Stabilizing the beam (active stabilization)

In Chapter 4 the most important lattice/optic parameters for machine commissioning were briefly discussed, and examples of their measurements were shown. Almost all of the previous described parameters are measured with the help of BPMs. More beam parameters and their (non-BPM based) measurements will be discussed in Chapters 6–9. This chapter discusses further BPM-based measurements and optimizations that are required mainly to stabilize the beam in the light source. Since the beam size in the IDs (e.g. undulators) is often smaller than 10 μm (vertical), the beam position at these locations must be as stable as 10% of the beam size. Larger position variations will strongly disturb photon beam experiments. The whole spectrum from short-term jitter on a bunch-by-bunch scale up to long-term drifts on a weekly scale should be avoided. In the following discussion it is assumed that the BPM system itself is already (more or less perfectly) stabilized within this required time range (see Chapter 2).

### 5.1    Orbit feed forward

Some orbit distortions cannot be avoided, even during user runs. Many facilities use off-axis injections by fast kickers to accumulate charges in selected bunches. The damping time of this injection oscillation is typically in the order of less than 1 ms due to damping by synchrotron radiation. These oscillations can easily be used to measure the damping time just by observing turn-by-turn the decaying oscillation amplitude of a single injected bunch [25]. During a small accumulation of new charge into a stored beam, a residual oscillation is, in general, small enough not to disturb the experiments too much. Typically, there is a relatively slow injection septum in the injection chain involved that separates the injection orbit from the stored orbit. The electromagnetic injection field might leak onto the area of the stored beam and might generate an additional orbit distortion of the stored beam over the time while the septum is powered (some milliseconds) [26]. Changing the gap of an undulator or other IDs can be another source of a slow orbit distortion [27].

These and other kinds of slow and predictable (!) distortions can be analyzed by the ORM method. Sets of correctors can be found and powered to cancel the orbit excursions, depending, for example, on the timing behaviour of the septum or the gap settings of the ID. Quite often a feed forward is realized by lookup tables (calculated by ORM) or by measured functional dependencies [28, 29], which are activated at the same time that the device is powered. This can reduce the orbit distortion to less than 10% of the uncorrected case.

## 5.2 Orbit feedback

As mentioned above, the beam stability, especially in the IDs, should be as good as one-tenth of the beam diameter. This requires in modern light sources a stability of better than 1 μm, mainly in the vertical plane. Diffraction-limited storage rings will need this stability in both planes. Good BPMs already deliver a resolution at the sub-μm level at a bandwidth of about 1 kHz (see Chapter 2), while acting as the primary input source of a global orbit feedback. X-ray BPMs from the beam lines are often also included in the feedback input; due to their long lever arm they are even more sensitive to orbit distortions [2]. The outputs of a global orbit feedback system are connected to the fast corrector magnets installed around the ring. These magnets and their power supplies have a typical bandwidth of up to 1 kHz and short-term noise figures of <1 ppm. There are typical sources of more or less unavoidable orbit distortions at various frequencies, e.g. influence from the mains electrical supply (50 or 60 Hz), vibrations stimulated by culture noise driving resonant vibrations of girders and supports (5–20 Hz), electrical stray fields, including cycling booster (1–>100 Hz), thermal effects, including cooling water (<1 Hz), ground settling and seasonal changes (<<0.1 Hz). Beam instabilities can drive orbit oscillations of far beyond 1 kHz; these have to be addressed by multi-bunch feedback systems (see Chapter 5.3). Due to the limited bandwidth of the corrector magnets and the internal feedback loop latency (>30 μs) the overall bandwidth of a modern fast orbit feedback typically does not exceed 1.5 kHz [30]; a bandwidth of 200 Hz is usually sufficient to counteract the main sources of orbit oscillations [30, 31].

The correction algorithm is typically based on the inversion of the ORM, which relates the beam position at the location of the BPMs with the corrector magnets kicks, using the technique of singular value decomposition (SVD). A direct acquirement of turn-by-turn beam position data from the BPM electronics will minimize the latency of the feedback loop. Fast data links between the BPM electronics and the (FPGA- or DSP-based) feedback electronics are required, as well as between the corrector magnets power supplies and the feedback output. Further examples of modern realizations of orbit feedbacks can be found in Refs. [32, 33].

## 5.3 Multi-bunch instabilities and feedback

Almost all high current synchrotron light sources require a fast multi-bunch feedback system to counteract coupled bunch instabilities that are driven by the resistive wall due to the small gaps of the undulators, wakefields, impedances, ion trapping and more. The threshold of these instabilities is at about some 10 mA. References [34, 35] gave very comprehensive discussions of the various collective effects. The performance of a feedback system should be adequate to suppress the fastest growth rate of an instability.

A multi-bunch feedback system detects an instability using one or more BPMs and acts back on the beam to damp the oscillation through a fast kicker. A dedicated BPM takes either the transversal position information or the longitudinal timing information on a bunch-by-bunch rate and sends this to the adjacent transversal or longitudinal feedback processor. A deviation from the zero position is then analyzed, the kicker strength is calculated to cancel this deviation, and the required signal is sent to the kicker at the required timing. A detailed discussion of bunch-by-bunch feedback systems can be found in Refs. [36, 37].

The effect of an instability can be seen in the transversal or longitudinal tune spectrum, where many additional lines appear. Transversal beam oscillations are visible in any BPM with bunch-by-bunch readout capability. But a bunch-by-bunch beam size measurement is also able to detect instabilities, since the transverse oscillation causes large off-centre positions in the quadrupoles, which causes beam size growth, even along a stored bunch train [38, 39].

The typical instrument to observe longitudinal instabilities is a streak camera. A streak camera shows turn-by-turn the longitudinal dimension of a bunch or a train of bunches. Longitudinal oscillations are therefore clearly visible [40].

Beam instabilities depend usually on the chromaticity of the beam. High chromaticity is necessary for suppressing various kinds of beam instabilities, while low chromaticity is good for high injection efficiency (larger dynamic aperture). A scan of the chromaticity versus an instability monitor is useful for finding a good compromise of chromaticity settings. Often a scan of different beam and bunch currents at various chromaticities as well as at different bunch repetition rates or ID gaps is very helpful for diagnosing and understanding the sources of instabilities [41, 42].

# 6      Beam current

Beam current monitors are essential for every kind of particle accelerator. Current transformers are widely used for non-destructive current measurements in storage rings [43]. Special DC current transformers (DCCTs) are in use to determine very precisely the overall current stored in the synchrotron with a resolution of a few microamperes. Fast current transformers (FCTs) are in use to measure the charge stored in individual bunches. Their bandwidth is limited to about ≤1 GHz, therefore they cannot be used at higher bunch repetition rates. A cross-calibration between the measured DC current and the sum of all individual bunch charges is most helpful to ensure a valid beam current measurement in a storage ring. Reference [44] uses the beam current output (sum of all four inputs) of the BPMs to crosscheck the DCCT readings. During (re-) commissioning of a storage ring, a very important feature is to look at, at least turn-by-turn, the sum of all four electrodes of each BPM's electronics. Since they are distributed around the ring their sum signal versus their position in the ring shows how far the beam travels before the first turns appear. At the location of a (full) beam loss the operator can search for faulty settings (e.g. orbit, magnet polarity, etc.) and can correct them to guide the beam somewhat further downstream.

## 6.1     Injection: Top-up, filling pattern, transfer efficiency

At moderate bunch repetition rates an FCT can be used to determine the charge in every bunch of the stored bunch train (filling pattern), together with all extended gaps between the bunches. Figure 8 shows a raw signal of an FCT with minimum bunch distance of 8 ns as well as some gaps between trains of bunches. Since a transformer has an AC coupling, the drop of the signal is clearly visible. Therefore, a sampling at the bottom line and on top of each individual bunch signal is necessary to acquire the charge of each individual bunch. With such bunch pattern measurement thresholds for topping-up, the charge of an individual bunch can be set. By reaching the minimum threshold a defined number of charges can be refilled on top of the bunch to restore the charge up to the maximum threshold. This allows keeping a homogenous filling pattern of better than a few percent (see Fig. 9) or, if required, special bunch pattern fillings with different charges and pilot bunches.

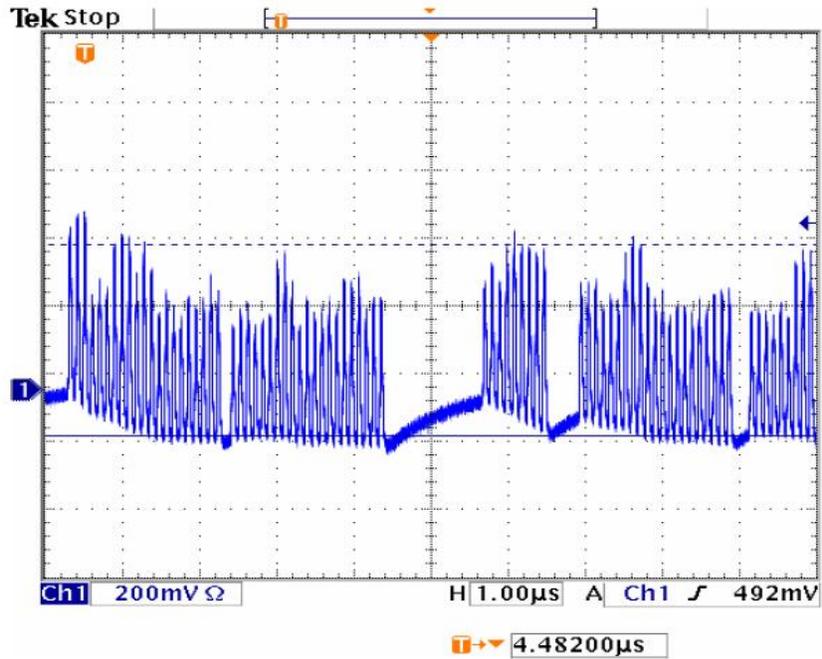

**Fig. 8:** Raw signal of an FCT at PETRA III with minimum bunch distance of 8 ns and various gaps between trains of bunches measured by a high bandwidth oscilloscope.

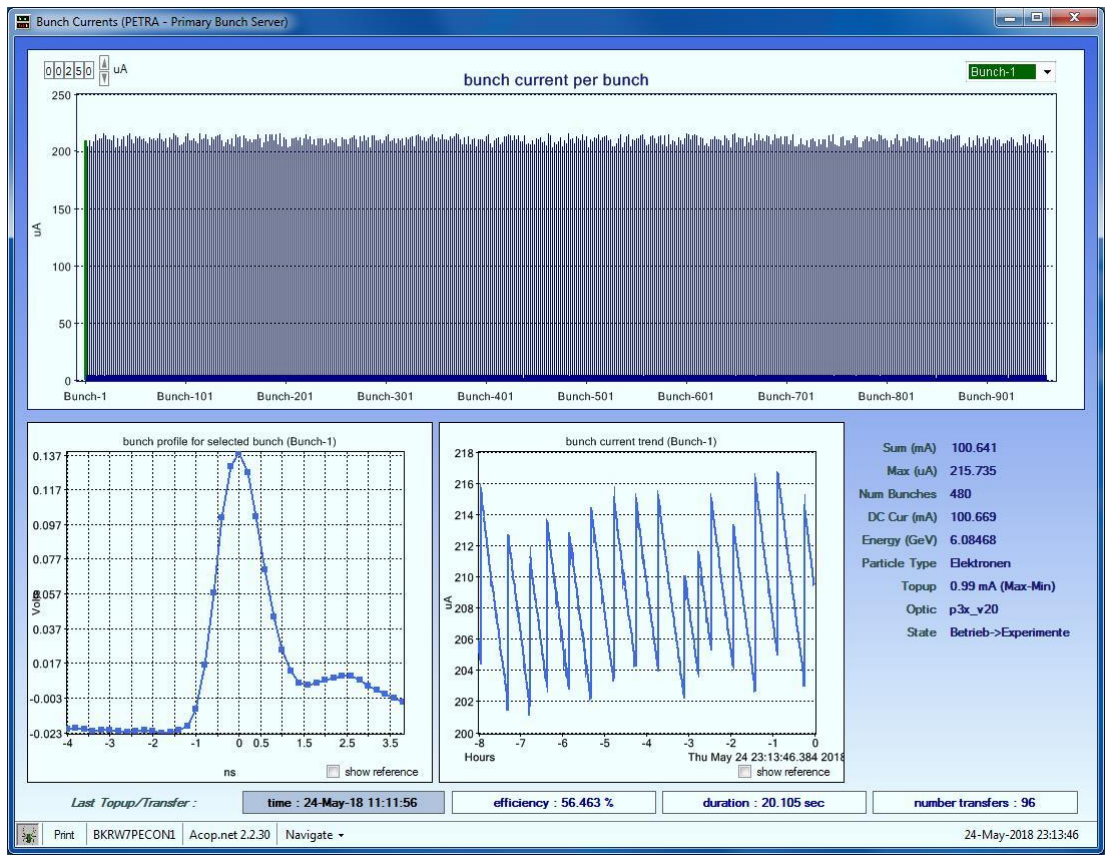

**Fig. 9:** Top: Constant filling pattern of 480 bunches in PETRA III. Lower left: Raw signal of a single bunch measured by an FCT together with all sampling points. Lower right: Current history of the first bunch over some hours. The top-up of the current starts at <204 µA. From Ref. [5].

To measure the filling pattern of the machine various methods are in use for light sources, e.g. the sum of all channels of a BPM, the amount of synchrotron radiation of each bunch, and time-correlated single photon counting (see Chapter 8). Reference [45] shows a nice comparison of various methods and an example of the top-up refill calculation procedure.

Once an individual bunch measurement is established in the whole (pre-)accelerator chain of a synchrotron light source, a comparison of the charge before and after injection into the synchrotron defines immediately the injection efficiency. Since there are not just fast losses, one has to define a certain time delay before measuring the stored beam. Typical delay times are defined by the tunes (transversal, longitudinal) of the synchrotron. Such a display of efficiency is a very nice tool for the operators to optimize the injection and lower beam losses and therefore lower the activation of critical machine parts.

## 6.2 Lifetime, dynamic aperture

From the decay of the stored charge over time ($N(t)$) the lifetime $\tau$ can be calculated by

$$N(t) = N_0 \times e^{\frac{-t}{\tau}} .$$

The lifetime in an electron storage ring is usually measured by a precise DCCT. The lifetime is determined by many effects, e.g. quantum excitation ($\tau_q$), elastic ($\tau_{el}$) and inelastic scattering ($\tau_{inel}$) on the residual gas atoms, scattering of electrons within the bunch (Touschek-effect; intra-beam scattering, IBS) ($\tau_{tou}$, $\tau_{IBS}$, respectively), and trapping of charged particles in the beam potential ($\tau_{ion}$). Electrons that interact by one of these effects receive a transverse or longitudinal kick, and their transversal (betatron) amplitude or longitudinal (synchrotron) amplitude increases. They might reach amplitudes where nonlinear effects start to dominate, and additional beam losses occur. This region is called 'dynamic aperture', transversal and longitudinal. A simple way to measure the transversal dynamic aperture or dynamic acceptance is to move a calibrated scraper into the beam. At a position $x_a$ where the lifetime gets shorter, the linear acceptance $A$ (geometric aperture) $A = x_{a2}/\beta$ is equal to the dynamic aperture of the machine [34]. During such a scraper scan a careful observation of the tune and of instabilities is necessary. Figure 10 shows an example of such a scan.

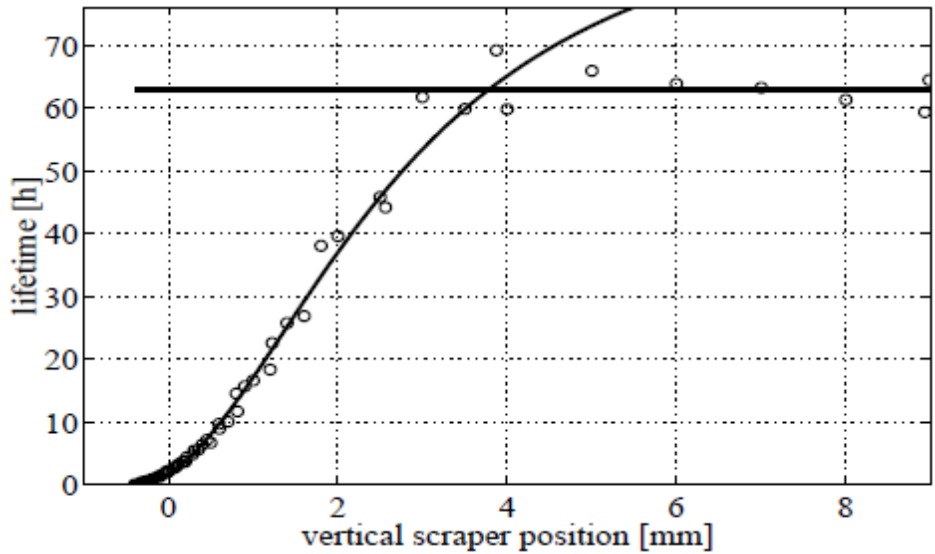

**Fig. 10:** Lifetime as a function of the vertical scraper position. The scraper position defines the geometric aperture. At the crossing point the dynamic aperture is the same as the geometric aperture. The solid lines are fits to the data. From Ref. [46].

Additional information regarding the source of limited lifetimes in a storage ring can be gained by measuring the lifetime at the same total current but with varying bunch currents. The total current keeps the gas pressure constant (constant scattering on residual gas atoms) while other effects like IBS and the Touschek effect depend on the bunch charge. Additional scans of the RF voltage versus lifetime show the dependence of the Touchek effect upon the momentum aperture. With all these lifetime measurements one can fit the various parameters of the different lifetime formulas to the measurements, receiving the lifetime contributions of all the effects to the total lifetime $\tau$ of the stored beam. Many of these measurements are discussed in detail in Refs. [46, 47].

## 7 Beam size: Emittance

The vertical emittance in third-generation synchrotron light sources is usually quite small: modern machines try to reach $\varepsilon_v = 1$ pm, while the horizontal emittance $\varepsilon_h$ is about a factor 100–1000 larger (depending on the coupling, see also Table 1). The emittance is derived from non-destructive beam size measurements by knowing the optical parameters (horizontal and vertical β-functions $β_{v,h}$; dispersion $D_{v,h}$) at the profile monitor

$$\varepsilon_v^2 = \frac{\sigma_v^2}{\beta_v + \left(D_v \times \frac{\Delta p}{p}\right)^2} \ , \ \varepsilon_h^2 = \frac{\sigma_h^2}{\beta_h + \left(D_h \times \frac{\Delta p}{p}\right)^2} \ ,$$

where $\Delta p/p$ is the relative momentum spread. The vertical dispersion $D_v$ is usually negligible. The two beam sizes are significant different, however, so that the beam profile monitor has to cover a large aperture while keeping its resolution in the range ≤1 μm. Synchrotron radiation diagnostics are the main tool for non-destructive size measurement in synchrotron light sources [48]. X-ray pin-hole cameras or double-slit interferometers are state-of-the-art instruments in third-generation light sources, see also Ref. [49]. Figure 11 shows an example of such a high resolution beam size measurement and emittance determination in PETRA III by a 2D interferometer. The coupling between the vertical and horizontal plane can be derived from the quotient of the two measured emittances [50].

A bunch-by-bunch beam size measurement enables a fast detection and diagnostic of instabilities by plotting each bunch size along the bunch train, e.g. at different chromaticities or bunch currents, see Ref. [39].

A turn-by-turn profile measurement is very useful to observe beam shape oscillations, e.g. at injection. These quadrupole oscillations result from an optic mismatch of the incoming beam and the storage ring lattice, and cannot be seen by a BPM system. An example of such a measurement is shown in Ref. [51].

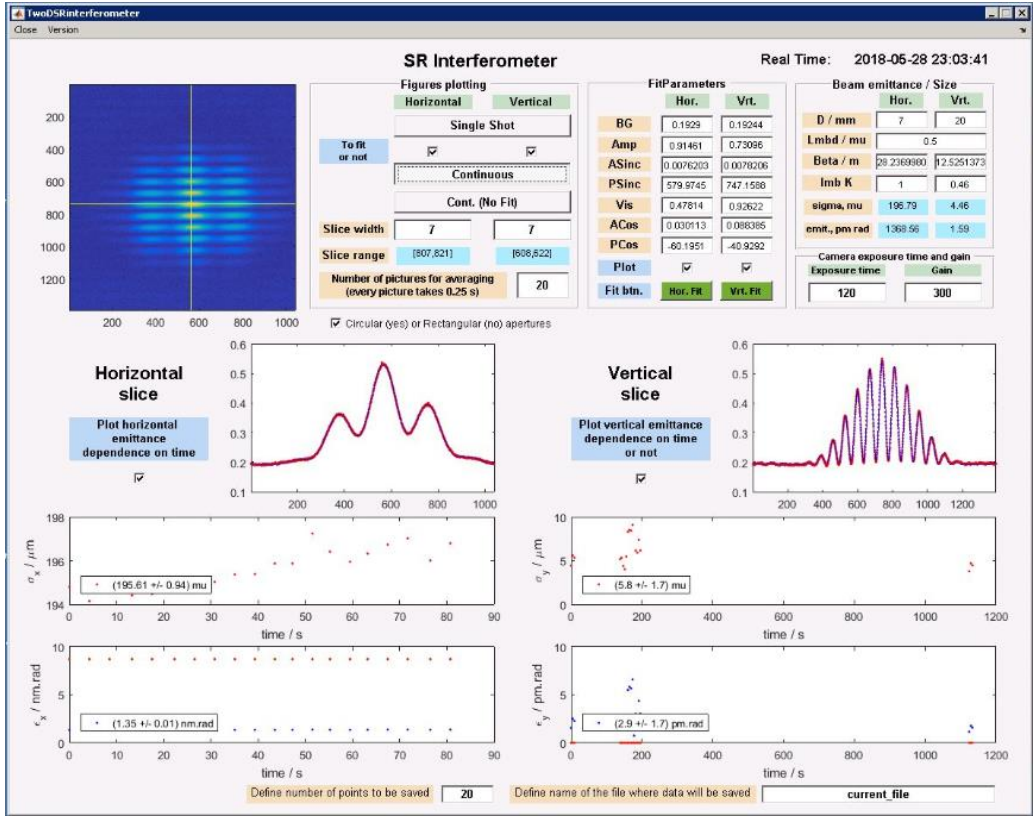

**Fig. 11:** Emittance measurement with a 2D interferometer in PETRA III. The measured beam sizes are $\sigma_v$ = 5.8 μm, $\sigma_h$ = 196 μm, the corresponding emittances are $\varepsilon_v$ = 1.59 pm rad, $\varepsilon_h$ = 1368.56 pm rad. from Ref. [5].

## 8   Bunch length: Energy spread, instabilities, bunch purity

The bunch length $\sigma_l$ in a synchrotron is a result of the momentum compaction factor $\alpha$ of the lattice, the synchrotron frequency $f_s$ and the energy spread $\Delta p/p$ in the machine

$$\sigma_l = \frac{\alpha}{2\pi f_s} \times \frac{\Delta p}{p} \ .$$

Bunch lengths between 20–40 ps and an energy spread of about $10^{-3}$ are typical values for synchrotron light sources. Optical streak cameras with a resolution of down to 0.5 ps are state-of-the-art instruments for measuring the bunch length with the help of optical synchrotron radiation. The smallest resolution is required for so-called 'low alpha lattice operation' in some machines [52]. If the bunch length is short enough, coherent synchrotron radiation is emitted and can be used for bunch length measurements as well [53]. A streak camera is fast enough to measure the bunch length and the arrival time of many bunches on a bunch-by-bunch rate. Therefore, it can also be used to prove the efficiency of the multi-bunch feedback system. At stable beam conditions the distance between all bunches is constant while it is changing during a longitudinal multi-bunch instability [54, 55]; even the mode of the instability can be determined by analyzing the time variation structure.

Since the bunch length depends on the synchrotron frequency $f_s$, one can use this to determine the momentum compaction factor $\alpha$ of the lattice by changing $f_s$ (varying the RF gap voltage) while observing the bunch length. Examples of using such diagnostic methods for determining $\alpha$ can be found in Refs. [56, 57]. By knowing $\alpha$ and measuring $f_s$ and the bunch length $\sigma_l$ one can calculate from the formula above the energy spread $\Delta p/p$ [58].

Zotter's bunch length scaling law [59] describes a relation between the bunch length, its charge (bunch current) and the effective impedance $\text{Im}[(Z/n)_{\text{eff}}]$ of the ring. By measuring the bunch length versus the bunch current, a fit to this curve while applying the scaling law delivers the impedance and an estimate of the ring's broad band inductance L, where $L\omega_{\text{rev}} = \text{Im}[(Z/n)_{\text{eff}}]$ and $\omega_{\text{rev}}$ is the revolution frequency.

Any disagreement of such a fit shows immediately further effects on the bunch lengthening like, for example, IBS. Therefore, this analysis should always be done below any instability threshold. More details and measurements can be found in Refs. [60, 61].

## 8.1 Bunch purity

In the case of a short bunch distance the bunch current measurement might not have a high enough bandwidth to measure precisely the filling pattern in the machine. An alternative is the time-correlated single photon counting method (TCSPC), which delivers a high dynamic range filling measurement of all(!) the possible RF buckets of the ring. It provides a timing resolution in the sub-nanosecond regime. The arrival time of single synchrotron radiation photons emitted by the electron bunches passing through a particular dipole in the storage ring is measured. The arrival time is measured relative to a clock pulse that is synchronized to the bunch revolution frequency via the storage ring RF system. The amplified signal is analyzed using a time-to-digital-converter (TDC) and a multi-channel analyzer (MCA). Commercial devices with high resolution and low dead time are available. A proper choice of the detector, e.g. Micro Channel Plate – Photo Multiplier (MCP-PMT), Avalanche Photodiode (APD), is necessary [62, 63]. A disadvantage of this method is the time required; many minutes of counting are required to get a dynamic range of $10^6$. Many more details of TCSPC are discussed in Ref. [64].

A requirement of some experiments at the beam lines in synchrotron light sources is a bunch purity of neighbouring bunches (buckets) up to $10^{-8}$. Quantum lifetime, residual gas scattering, Touschek scattering and injection errors are the main mechanisms of filling neighbouring buckets [65]. For example, Touschek scattering can lead to a jump of electrons from a highly charged bucket into one of the neighbouring buckets, resulting in a continuous filling of the neighbouring buckets [66]. Therefore, an active bunch cleaning is required to satisfy the requests of the experiments. Since the transversal tune of a highly charged bunch is slightly different than that of a low-charged bunch (a few kHz), an excitation by a transversal kicker on just the low charge tune frequency will excite large oscillations and losses of this bunch while leaving the highly charged bunch almost unaffected. This cleaning effect can well be detected and followed by the TCSPC method [66, 67].

## 9  Beam losses: Radiation damage to undulators

Many beam diagnostic experiments can be done with the help of beam loss monitors (BLMs) [68]. These are very useful during the commissioning of a light source [69] as well as for improvements of the injection efficiencies [70], although the top-up mode in many machines can easily cope with bad injection efficiencies.

Beam losses, however, create additional radioactive radiation and serious radiation damage to sensitive components like IDs [71]. The geometric aperture of a ring is often set by absorbers (collimators) somewhere in the ring in a way that they are just a little bit closer to the beam than the IDs. Therefore, the sensitive IDs should somehow be in the shadow of the absorbers. This can be proven by observing the loss rate with BLMs located at the IDs while slowly opening the absorbers. Under 'normal' conditions the loss rate at the IDs is constantly low as long as the IDs are in the shadow; their loss rate increases drastically as soon as the absorbers no longer give a shadow, and the geometric aperture is then defined by the IDs. Under conditions where the lifetime of the beam is limited by the Touschek effect the absorbers (not necessarily near the ID) will not produce a clear shadow, since Touschek losses can appear at every location in the ring. Therefore, each step when opening an absorber will produce an increase of the loss rate at the IDs [71]. With the help of BLMs a proper setup of the absorbers can be found to minimize the radiation damage to the IDs by leaving enough apertures to achieve a sufficient lifetime of the stored beam.

## 10    Summary and outlook

Many beam diagnostic examples in third-generation synchrotron light sources have been given. Most have been touched upon very briefly, while more information can be found in the extensive reference list.

One of the main instruments for beam diagnostics are BPMs, which are used to determine not only the beam position and beam orbit, but also to diagnose many beam optic parameters like $\beta$-function, tune, dispersion and chromaticity etc. Of huge importance in any third-generation synchrotron light source is the beam stability. This includes mechanical vibrations, thermal drifts of electronics and components, and orbit stability. Some examples related to this topic have been given. The help from various feedback systems to stabilize the beam was also discussed.

Examples of the use of beam current monitors to diagnose filling patterns in the machine, transfer efficiencies, lifetimes and dynamic aperture were given. Beam size monitors are used to diagnose the transversal emittance of the beam while bunch length monitors are used to measure the longitudinal emittance, energy spread and bunch purity, but both are also used to observe beam instabilities. Last but not least an example of the use of BLMs was given to protect the IDs from radiation damage.

For the user operation of a future diffraction limited light source (not discussed in this report), the pointing stability of the synchrotron radiation is one of the most important issues. Like today, the stability target of the optical axis of each photon beamline should be as good as one-tenth of the beam size, corresponding to <0.5 μm for the source position, but in both planes. The angular stability should be smaller than one-tenth of the beam divergence resulting in <0.5 μrad for the angle, also in both planes. Reference [72] discusses some more important features of beam instrumentation and diagnostics for these kinds of new machine.

**Acknowledgements**

Many thanks to my colleagues Gero Kube from DESY for adding a lot of useful hints and comments to this manuscript, and to Joachim Keil, DESY, for discussing with me a lot of beam dynamics experiments at PETRA III.